\documentclass[preprint,12pt]{elsarticle}

\usepackage{amssymb,setspace}
\usepackage{color}
\usepackage{multirow}
\usepackage{subcaption}
\usepackage{tcolorbox}
\usepackage{tabularray}

\begin{document}

\begin{frontmatter}

\author[inst1]{Chuanbo Hu}
\affiliation[inst1]{organization={Department of Computer Science, University at Albany},%Department and Organization
            addressline={1400 Washington Ave}, 
            city={Albany},
            postcode={12222}, 
            state={NY},
            country={USA}}

\author[inst2]{Bin Liu}
\affiliation[inst2]{organization={Department of Management Information System, West Virginia University},%Department and Organization
            addressline={83 Beechurst Avenue}, 
            city={Morgantown},
            postcode={26506-6025}, 
            state={WV},
            country={USA}}

\author[inst3]{Minglei Yin}
\affiliation[inst3]{organization={Lane Department of Computer Science and Electrical Engineering, West Virginia University},
            addressline={1220 Evansdale Drive}, 
            city={Morgantown},
            postcode={26506-6025}, 
            state={WV},
            country={USA}}

\author[inst4]{Yilu Zhou}
\affiliation[inst4]{organization={Department of Information, Technology, and Operations, Fordham University},%Department and Organization
            addressline={140 W. 62nd Street}, 
            city={New York, NY},
            postcode={10023}, 
            state={NY},
            country={USA}}

\author[inst1]{Xin Li}



%\title{Can ChatGPT Protect Children from Age-Inappropriate Apps? A Study on Maturity Rating with Multimodal Large Language Models}
% other title options \titie{:}
\title{Multimodal Chain-of-Thought Reasoning via ChatGPT to Protect Children from Age-Inappropriate Apps}


\begin{highlights}

\item We conduct the first systematic study on app maturity rating with multimodal large language models (MLLMs). 
\item We design chain-of-thought (CoT) endowed prompting to instruct MLLM to follow a sequence of logical steps to derive maturity levels for apps. 
\item We conduct extensive experiments on datasets collected from the App Store with different LLM models and different multimodal fusion strategies, and the results demonstrate the effectiveness of the proposed method.

%\item Practical Implications for Digital Safety: Offers insights into how advanced AI tools can be applied in real-world scenarios to enhance digital safety and inform better governance of digital content.

\end{highlights}

\begin{abstract}
Mobile applications (Apps) could expose children to inappropriate themes such as sexual content, violence, and drug use.
\emph{Maturity rating} offers a quick and effective method for potential users, particularly guardians, to assess the maturity levels of apps.
Determining accurate maturity ratings for mobile apps is essential to protect children's health in today's saturated digital marketplace.
Existing approaches to maturity rating are either inaccurate (e.g., self-reported rating by developers) or costly (e.g., manual examination). 
In the literature, there are few text-mining-based approaches to maturity rating.  However,  each app typically involves multiple modalities, namely app
description in the text, and screenshots in the image. 
In this paper, we present a framework for determining app maturity levels that utilize multimodal large language models (MLLMs), specifically ChatGPT-4 Vision. 
Powered by Chain-of-Thought (CoT) reasoning, our framework systematically leverages ChatGPT-4 to process multimodal app data (i.e., textual descriptions and screenshots) and guide the MLLM model through a step-by-step reasoning pathway from initial content analysis to final maturity rating determination. As a result, through explicitly incorporating CoT reasoning, our framework enables ChatGPT to understand better and apply maturity policies to facilitate maturity rating. 
Experimental results indicate that the proposed method outperforms all baseline models and other fusion strategies.

%This improved predictive ability not only offers guardians a dependable maturity reference but also aids in the broader societal effort to protect young individuals from the risks associated with digital interactions. 

\end{abstract}



\end{frontmatter}


\section{Introduction}

The widespread use of mobile devices (e.g., smartphones and tablets) has reshaped our engagement with digital content, typically provided in the format of mobile apps, in our daily lives. As of 2024, a staggering number of apps are available, with approximately 3.95 million on Google Play \cite{google_play_stats_2024} and about 1.83 million on Apple App Store \cite{apple_store_stats_2024}. In particular, children are increasingly using these devices for both educational and recreational activities. 
Despite the benefits, the proliferation of age-inappropriate content poses serious risks for young children. Exposure to apps featuring mature content such as violence,  sexual content, gambling, or drug use can have detrimental effects on children's psychological and behavioral development. 
Research in developmental psychology underscores that young children, when exposed to such content, may not only mimic these behaviors but also suffer long-term developmental consequences \cite{anderson2001effects,hill2016media}. Therefore, ensuring that children interact only with age-appropriate apps is crucial for safeguarding their growth and well-being in an increasingly digital world.

To help mitigate these risks, mobile platforms like Apple and Google have implemented app \emph{maturity rating} systems, akin to those used in the video game and movie industries. These systems are designed to inform users about the content of apps and to suggest the appropriate age group for their use.
For example, App Store implements a maturity rating policy that includes four levels: \emph{4+, 9+, 12+}, and \emph{17+} \footnote{https://developer.apple.com/help/app-store-connect/reference/age-ratings/}. These categories are designed to guide users on the appropriate age group for each app, helping to safeguard younger users from content that may be unsuitable for their age group. In addition, App Store also provides detailed descriptions of the content that justify these ratings. 
Table \ref{tab:motavation_case} shows an example of mature contents and maturity levels of apps at App Store. 
This transparency helps users understand the nature of an app and the reasons behind its specific maturity classification. Similarly, Google Play employs an age-based maturity rating that determined by different territories \footnote{https://support.google.com/googleplay/answer/6209544?hl=en}.
However, unlike the ratings issued by the Motion Picture Association of America (MPAA) and the ESRB, which are based on evaluations by independent rating committees, app ratings are often based on developers' self-reported assessments. This self-reporting approach can lead to inaccuracies. Developers may underestimate the maturity level of their apps, either inadvertently or to broaden their market reach, leading to a misclassification of content \cite{chen2013app}. This situation has caused growing concern among parents and educators, who frequently encounter apps with maturity ratings that do not accurately reflect their content.

\begin{table}[t]
\centering
\small
\caption{Illustrative example of mature contents and maturity levels of apps at App Store. Apps are categorized into four maturity levels, i.e., 4+, 9+, 12+, and 17+. In addition to the maturity level, the App Store also identifies the detailed mature contents that make an app be rated as a specific maturity level. }
\label{tab:motavation_case}
\begin{tabular}{p{2.7cm}|p{4.2cm}|p{5.4cm}}
\hline
\textbf{Screenshots} & \textbf{Description} & \textbf{Maturity Rating \& Content } \\ %& \textbf{GT} & \textbf{P\_ours} & \textbf{P\_CF} & \textbf{P\_SF} \\
\hline
\raisebox{-\totalheight}{\includegraphics[width=0.21\textwidth]{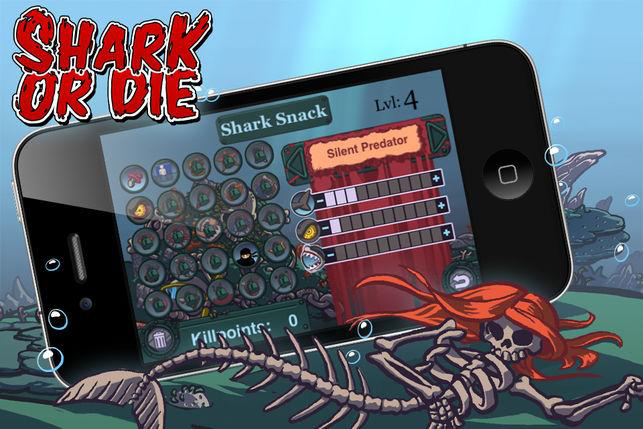}} {\includegraphics[width=0.21\textwidth]{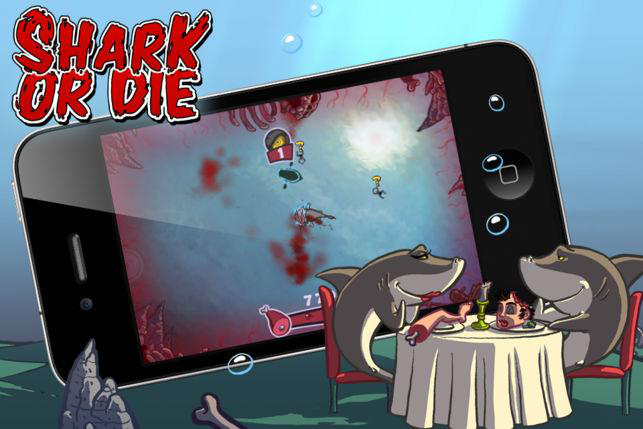}}  & Those nasty humans keep making shark fin soup, so it's time to turn the tables! Turn into an ocean predator; comes in the way of your mighty bite: swimmers, rubber ducks, beach babes, and more...   &     \textcolor{blue}{Rating: 17+}.  \textcolor{red}{Mature contents:} 
\textcircled{1} Frequent/Intense Cartoon or Fantasy Violence
\textcircled{2}  Frequent/Intense Realistic Violence
\textcircled{3}  Infrequent/Mild Sexual Content and Nudity
\textcircled{4}  Infrequent/Mild Horror/Fear Themes
\textcircled{5}  Frequent/Intense Profanity or Crude Humor
 \\ 
\hline
\end{tabular}
\vspace{-10pt}
\end{table}

%As illustrated in Table \ref{tab:motavation_case}, the diversity of maturity content ratings, from cartoon violence to realistic violence and sexual themes, underscores the necessity for robust parental oversight and stricter content regulation in digital platforms to safeguard young users. 

%Similarly, Google Play employs an age-based rating policy that varies by region. In the United States, the platform adheres to the Entertainment Software Rating Board (ESRB) guidelines, categorizing apps into Everyone', Everyone 10+', Teen', Mature', and `Adults Only' ratings \cite{esrb_ratings_guide}. In regions without a specific rating authority, Google Play uses the International Age Rating Coalition (IARC) framework \cite{esrb_iarc}. Google Play also details the specific content that informs each maturity level, aiding users in making informed decisions about their downloads.

%Reports have highlighted these discrepancies, emphasizing the need for a more reliable, objective, and automated rating system \cite{west2014limitations}. 

To the best of our knowledge, there has been limited research conducted on app maturity rating systems. Chen et al. \cite{chen2013app} studied maturity ratings across Android and iOS applications, identifying discrepancies and proposing methods to improve content safety for children. 
Hu et al. \cite{hu2015protecting} proposed a text classification based automatic maturity rating framework to access app maturity levels from app descriptions.
Zhou et al. \cite{zhou2022automatic} improved previous methodologies by integrating a more advanced machine learning model that uses both app metadata and user-generated content. However, previous studies on app maturity rating have  significant limitations. 
First, these approaches often fail to capture the full spectrum of content (both textual description and screenshot images) that can influence an app's maturity rating. 
Second, these approaches are based on conventional supervised machine learning methods, which typically required large amount of labeled data to train the models, and may not fully leverage the capabilities of more advanced analytical techniques.

Responding to these challenges, in this paper, we present a framework that utilizes multimodal large language models (MLLMs), specifically ChatGPT 4V,  to take input of multimodal data (i.e., textual descriptions and screenshots)  of apps to determine the maturity levels for apps.  Unique difficulties arise when applying the textual descriptions and screenshots information for app maturity rating. 
First, the main purpose of app description and its screenshots is to introduce its functionality and to promote the app on an app market (e.g, App Store) rather than to directly provide information regarding its maturity level. For example, the maturity information, if any, in the descriptions could be ambiguous and obscure. Meanwhile, unlike  maturity images in the film industry,  app screenshots reveal maturity content in a more complex way. 
Second, different screenshots would contain different levels of maturity contents.  As such, only a subset of the screenshots are useful in determining the maturity levels. 
To this end, we design chain-of-thought (CoT) endowed prompting for app maturity rating. 
CoT prompting enables ChatGPT-4V to follow a sequence of logical steps to derive maturity levels for apps. 
Specifically,  we first identify and rank the screenshots according to the \emph{exist} and \emph{intensity} of maturity contents for a given app according to the maturity rating policy. Accordingly, based on the policy, we design \emph{maturity content prompt} and \emph{maturity intensity prompt} to identify the maturity content and its intensity in each screenshot. Then we design a \emph{maturity rate prompt}, which combines the top screenshot(s) and the textual descriptions, to determine the final maturity rating for the app. 

We assess our proposed method through extensive experiments on a dataset that we collected from the App Store. The results show
that the proposed approach consistently outperforms baseline models by a significant margin and demonstrates the
effectiveness of the CoT-endowed prompting with MLLM for app maturity rating.

In summary, our key contributions are as follows:
\begin{itemize}
\item We conduct the first systematic study on app maturity rating with multimodal large language models (MLLMs). 
\item We design chain-of-thought (CoT) endowed prompting to instruct MLLM to follow a sequence of logical steps to derive maturity levels for apps. 
\item We conduct extensive experiments on datasets collected from the App Store with different LLM models and different multimodal fusion strategies, and the results demonstrate the effectiveness of the proposed method.
\end{itemize}

\section{Background and Literature Review}

\subsection{Mobile App Maturity Rating}
Apps could expose users to mature themes. Therefore, app platforms such as the App Store and Google Play implement maturity rating systems. 

\textbf{App Store's Maturity Rating Policy:} The App Store requires all published apps to submit maturity ratings according to its maturity rating policy. As outlined in Table \ref{Tab:ios_Rating_Policy}, the App Store categorizes apps into four maturity levels: \emph{4+}, \emph{9+}, \emph{12+}, and \emph{17+}. The categorization is based on the content's nature and intensity, such as violence, sexual themes, profanity, and substance use. For instance, an app may be rated \emph{12+} for mild violence but \emph{17+} if the violence is intense. The highest level of mature content detected determines the app's final maturity rating.

\begin{table}[t]
\small
\caption{ App Store's maturity rating policy  --- it examines the existence and intensity of mature contents, and assigns an maturity rating to each app. }
\label{Tab:ios_Rating_Policy}
{\begin{tabular}{  l  l  c  c}
\hline
    Maturity content (Reason)  & Intensity & Maturity rating & \# \\ \hline
      None & --- & 4+ & 1\\
    \hline
        \multirow{2}{*}{Cartoon or fantasy violence}
         & Infrequent/mild  & 9+ & 2\\
         &  Frequent/intense & 12+ & 3\\ \hline
    \multirow{2}{*}{Horror or fear themed content}
         & Infrequent/mild  & 9+ & 4\\
         &  Frequent/intense & 12+ & 5\\ \hline
    \multirow{2}{*}{Profanity or crude humor}
         & Infrequent/mild  & 9+ & 6\\
         & Frequent/intense & 12+ & 7\\ \hline
        \multirow{2}{*}{Mature or suggestive content}
        & Infrequent/mild & 9+ & 8\\ 
        & Frequent/intense & 17+ & 9\\ \hline
    \multirow{2}{*}{Sexual content or nudity}
         & Infrequent/mild  & 12+ & 10\\
         & Frequent/intense & 17+ & 11\\ \hline
    \multirow{2}{*}{Realistic violence}
         & Infrequent/mild  & 12+ & 12\\
         &  Frequent/intense & 17+ & 13\\ \hline
    \multirow{2}{*}{Alcohol, tobacco, or drug use}
         & Infrequent/mild  & 12+ & 14\\
         & Frequent/intense & 17+ & 15\\  \hline
    \multirow{2}{*}{Medical or treatment-focused content}
         & Infrequent/mild  & 12+ & 16\\
         & Frequent/intense & 17+ & 17\\  \hline
    \multirow{2}{*}{Simulated gambling}
         & Infrequent/mild  & 12+ & 18\\
         & Frequent/intense & 17+ & 19\\  \hline
    Gambling or contests & --- & 17+ & 20\\ \hline
    Unrestricted web access & --- & 17+ & 21\\ 
\hline
    \end{tabular}}
    %\hline
\vspace{-10pt}
\end{table}

\textbf{Google Play's Maturity Rating Policy:}
Google Play employs an age-based rating system, with maturity levels influenced by regional regulatory bodies \footnote{https://support.google.com/googleplay/answer/6209544?hl=en}. For example, in the United States, Google Play adheres to the Entertainment Software Rating Board (ESRB) standards, categorizing apps into ratings such as `Everyone', `Everyone 10+', `Teen', `Mature', and `Adults Only'. If no regional authority is present, ratings are assigned by the International Age Rating Coalition. 

%Google Play also details the types of content that contribute to each maturity rating, helping to clarify the reasons behind the assigned ratings.

%\textbf{Comparing the two policies:} While there are similarities and some equivalence between the two policies under certain conditions, notable differences exist in their rating scales and the methodologies used to assess the presence and intensity of mature themes.

%Mobile app platforms like the Apple Store and Google Play lack official rating organizations to assess app maturity levels. Instead, these platforms rely on developers' self-reported ratings via questionnaires about the content before publishing. The Apple Store employs a more stringent verification process to adjust misleading ratings, whereas Google Play allows apps to be downloaded shortly after submission. This reliance on self-reporting can lead to inaccuracies, exposing children to unsuitable content, and has prompted concerns among parents about the reliability of these ratings \citep{ftc,washingtonpost}.

\subsection{Text Mining Based App Maturity Rating}

The lack of accuracy and reliability in current self-reported app ratings underscores the need for an accurate automatic system for assessing app maturity. 
There has been limited research conducted on app maturity rating systems. Chen et al. \cite{chen2013app} proposed a keyword matching-based approach to rate app maturity, which often resulted in limited accuracy due to a lack of semantic understanding. 
Hu et al. \cite{hu2015protecting}  proposed a text classification-based maturity rating method, which combines an augmented keyword list and traditional bag-of-words models to access app maturity levels from app descriptions.
Zhou et al. \cite{zhou2022automatic}  proposed a machine learning-based method by integrating both app metadata and features of the apps.
Despite these advancements, previous methods predominantly focused on textual data from app descriptions and overlooked the rich information available in other modalities like screenshots. 
Our work distinguishes itself by integrating both text and image data, leveraging the complementary information from both to enhance maturity rating accuracy. 

%This holistic method not only increases the reliability of maturity ratings but also provides a more nuanced understanding of the content within apps, thereby setting a new standard in the field of app maturity rating systems.

%guo2024moderating
\subsection{Multimodal Large Language Models}
Multimodal learning has significantly advanced AI capabilities by effectively integrating diverse data types such as text, images, and audio and has led to success in various applications \cite{baltruvsaitis2018multimodal,ngiam2011multimodal,hu2021detection,hu2021identifying,hu2023fine,ramachandram2017deep}. 
Recently, the emergence of Large Language Models (LLMs) has further propelled this field, particularly through zero-shot learning abilities, where these models perform tasks they haven't been explicitly trained on, demonstrating remarkable flexibility and breadth of application \cite{wei2022emergent,chang2024survey,zhao2023survey,hu2023unveiling}. The introduction of specialized Multimodal LLMs, such as ChatGPT-4 Vision (GPT-4V), marks a pivotal development \cite{openai2023gpt4}. These models are not only adept at processing mixed data types (i.e., text and image) but also excel in generating coherent, contextually accurate outputs across different modalities, setting a new standard in AI's operational capabilities. Recent studies have begun to demonstrate the potential of GPT-4V in tackling domain-specific challenges across various fields. For example, Schramm et al. explored the impact of multimodal prompts in brain MRI diagnostics, demonstrating the efficacy of GPT-4V in analyzing complex medical imagery \cite{schramm2024impact}. Similarly, Pillai et al. evaluated the capabilities of GPT-4V in dermatology, particularly for diagnostic and treatment recommendations, marking a significant advancement in medical AI applications \cite{pillai2024evaluating}. Furthermore, Jia et al. investigated the use of GPT-4V in media forensics, specifically its ability to detect deepfakes, revealing its utility in digital content verification \cite{jia2024can}. Lastly, Lian et al. introduced a zero-shot benchmark for emotion recognition using GPT-4V, expanding its uses into emotional AI interactions \cite{lian2024gpt}. These studies collectively highlight the versatility of GPT-4V and its transformative impact across different fields. However, despite its advancements, ChatGPT still faces significant challenges, particularly in sophisticated reasoning and contextual understanding. It sometimes struggles with complex problem-solving that requires nuanced, logical deduction, highlighting a critical area for further development to enhance its effectiveness and reliability \cite{ahn2024large,wadhawan2024contextual}.

\subsection{Chain-of-Thought Reasoning Based on Large Language Model}
The Chain-of-Thought (CoT) reasoning approach has significantly enhanced the capabilities of LLMs, enabling them to tackle complex, multi-step problems by mimicking human-like reasoning processes \cite{wei2022chain}. Kojima et al. \cite{kojima2022large} demonstrate how these models excel in zero-shot reasoning tasks, advancing the field significantly. A self-consistency method has been proposed that refines CoT reasoning, improving the analytical accuracy of these models \cite{wang2022self}. Zhou et al. \cite{zhou2022least} further expand on this by exploring least-to-most prompting, enhancing complex reasoning in LLMs. Despite these advancements, the integration of CoT with multimodal models, which process both text and visual data, remains largely untapped, promising significant potential for areas requiring nuanced analysis, such as app maturity rating.

\section{Methods}

\subsection{Problem Formulation}
\label{sec:problem_formulation}

Focusing on App Store,  we aim to design an effective framework that leverages the multimodal data associated with apps, namely app descriptions and screenshot images, to predict mature ratings for the apps under the guideline of the maturity rating policy of an app market (e.g., App Store). Specifically,  for each app $i$, we have access to its textual \emph{description} $D_{i}$, and its $n$ \emph{screenshot images}   $S_{i} =[s_{1},s_{2}, ..., s_{n}]$. We also have knowledge of the \emph{maturity rating policy} as shown in Table \ref{Tab:ios_Rating_Policy}. The policy establishes a direct correlation between the Maturity Content (MC) indicators and their corresponding maturity ratings ($R$), where the intensity ($I$) of each $MC$ influences the level of maturity rating assigned. For each app $i$, there are a set of 12 $MC$ indicators $MC_{i}=[MC_{1}, MC_{2}, ..., MC_{12}]$, and its maturity rating  $R_i\in \{R_{1}, R_{2}, R_{3}, R_{4}\}$ which is aligned with four age groups: $4+, 9+, 12+$, and $17+$.
Our goal is to design an effective system to determine the maturity levels for apps.

\subsection{Overview of the Proposed Framework}

\begin{figure}[t]
\centering
\includegraphics[width=1\linewidth]{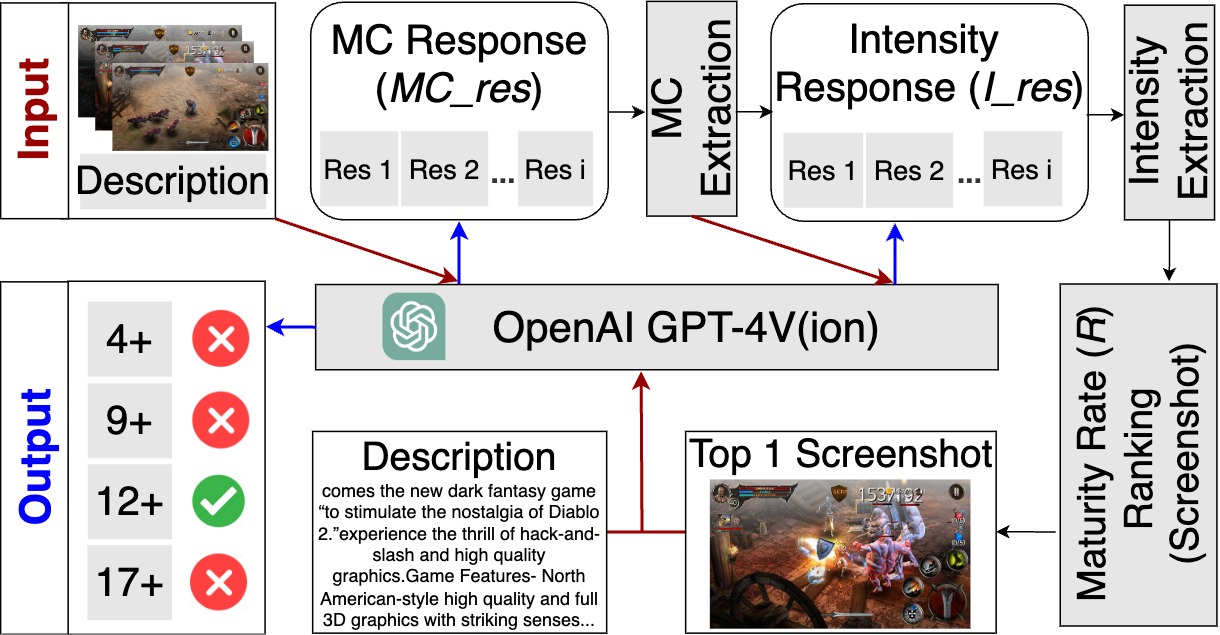}
\vspace{-10pt}
\caption{ Illustration of the proposed framework for app maturity rating using multimodal large language model (GPT-4V) with chain-of-thought (CoT) reasoning. The inputs (app descriptions and screenshots)  are denoted in \textcolor{red}{red} font, while outputs (maturity ratings) are shown in \textcolor{blue}{blue} font. Requests to GPT-4V are denoted with \textcolor{red}{red} arrows, and responses from GPT-4V are marked with \textcolor{blue}{blue} arrows.
Maturity content and intensity extraction aims to identify and rank the screenshots according to the exist
and intensity of maturity contents. 
The final maturity rating is 
determined by combing the top screenshot(s) and the textual descriptions.}
\label{fig:structure}
%\vspace{-10pt}
\end{figure}

Building on aforementioned problem formulation, we propose a framework that utilizes multimodal large language models (MLLMs), specifically ChatGPT-4 Vision, to determine the maturity levels for apps. As shown in Figure \ref{fig:structure}, our framework takes the multimodal input including textual descriptions $D$ and screenshots $S=[s_{1},s_{2}, ..., s_{n}]$ of apps, which are crucial in reflecting the app's content under the specific rating policy of an app platform, for ChatGPT-4V to access maturity levels.
On the one hand, the maturity information in the descriptions could be ambiguous and obscure.  On the other hand, different screenshots would contain different levels of maturity contents, and only a subset of the screenshots are useful in determining the maturity levels. 
Therefore,  we further endow our framework with chain-of-thought (CoT) reasoning, with which the ChatGPT-4V follows a sequence of logical steps to derive maturity levels for apps. The detailed steps are summarized as follows:

{\noindent \textbf{Step 1: Maturity Content and Intensity Extraction.}}
Given the screenshots $S=[s_{1},s_{2}, ..., s_{n}]$ of an app, we utilize CoT reasoning to enable GPT-4V to process each screenshot sequentially to deduce the maturity content $MC$ and its intensity levels $I$. 
Since different screenshots would contain different levels of maturity contents,  this step-by-step reasoning will identify and rank the screenshots according to the \emph{exist} and \emph{intensity} of maturity contents in the given app according to the maturity rating policy. 

%\item \textbf{Maturity Content and Intensity Evaluation:} The model employs CoT based on GPT-4V to dissect the mature content indicators $MC$ and $I$ for each screenshot $S$. This step helps to understand the screenshots related to the degree of maturity of content.

{\noindent \textbf{Step 2: Maturity Rating Determination.}} The maturity ratings $R$ are computed by combing the top screenshot(s) $S^{\ast} \in \{s_{1},s_{2}, ..., s_{n}\}$ identified in step 1 and the textual descriptions $D$.

%Through this methodology, the framework capitalizes on GPT-4V natural language understanding and image processing capabilities to comprehensively assess app content. This systematic approach allows for a detailed evaluation, supporting guardians and users in making informed decisions about app downloads based on their maturity ratings.

\subsection{Chain-of-Thought (CoT)  Endowed Prompting for App Maturity Rating}

\begin{figure}[t]
\centering
\includegraphics[width=1\linewidth]{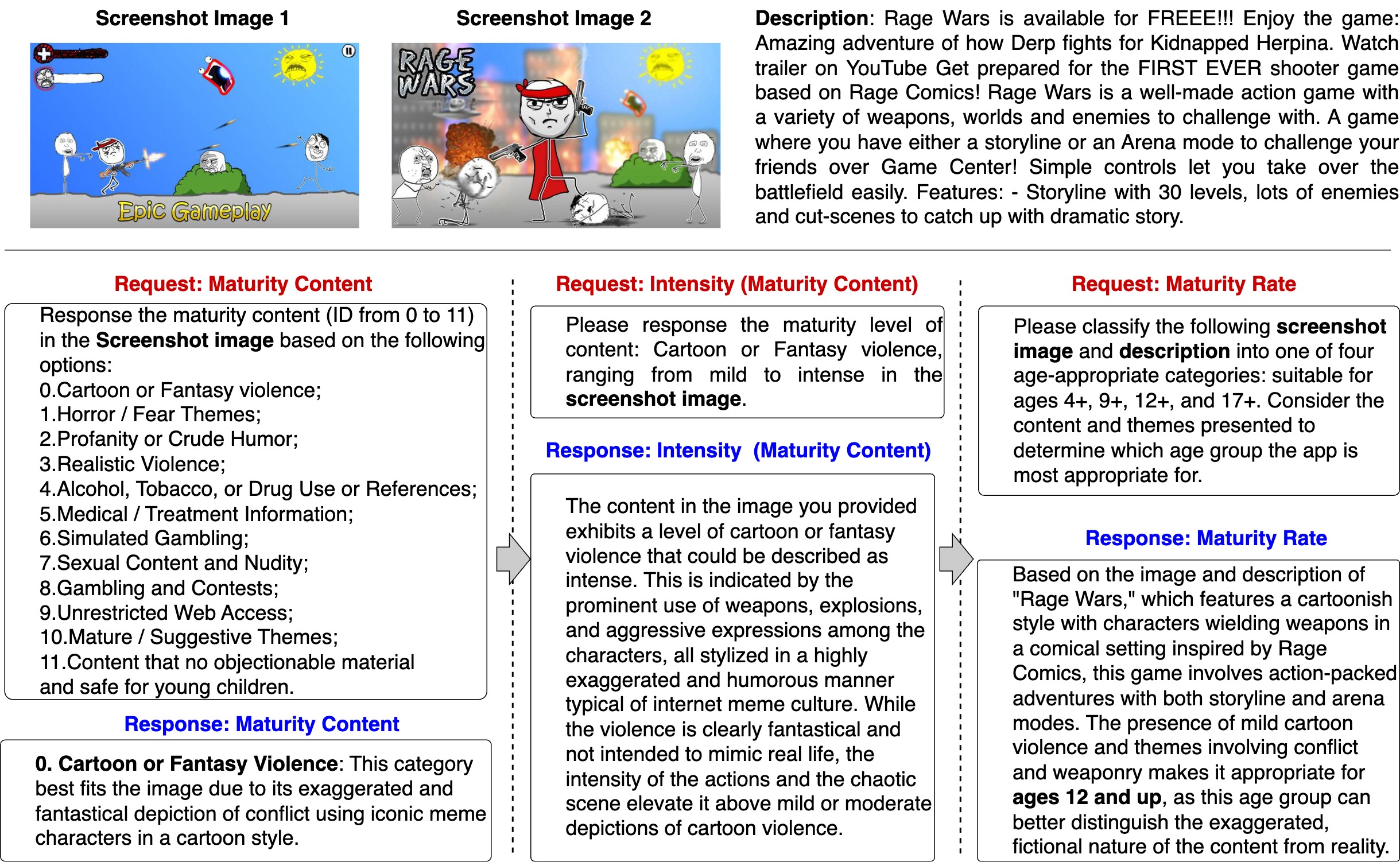}
\vspace{-5pt}
(a) ~~~~~~~~~~~~~~~~~~~~~~~~~~~~~~(b) ~~~~~~~~~~~~~~~~~~~~~~~~~~~~~~(c)\\
\caption{Illustration of chain-of-thought (CoT) reasoning endowed prompting for app maturity rating using an app with two screenshots and a description as an example. The intermediate  steps in the CoT reasoning include (a)  maturity content prompt and (b) maturity intensity prompt to identify
and rank the screenshots according to the maturity rating policy. Finally, (c) maturity rating prompt  combines screenshot(s) and textual
description to generate a maturity rating score. 
Requests to GPT-4V are highlighted in \textcolor{red}{red} font, and responses are in \textcolor{blue}{blue}. }
\label{fig:cot_prompt}
%\vspace{-10pt}
\end{figure}

The effectiveness of using a MLLM for app maturity rating relies heavily on the design of specialized prompts that analyze both textual description $D$ and screenshot images   $S =[s_{1},s_{2}, ..., s_{n}]$ with CoT reasoning. 
As shown in Figure \ref{fig:cot_prompt}, our CoT-endowed prompting enables maturity rating  reasoning  through a sequence of  reasoning steps, and it is composed  of 
\emph{maturity content prompt} and \emph{maturity intensity prompt} to identify the maturity content and its  intensity in each screenshot, and a \emph{maturity rate prompt}, which combines the top screenshot(s) and the textual
descriptions to determine the final maturity rating for the app. 

%\subsubsection{Maturity Rating Policy based Prompting}

\textbf{Maturity Content Prompt:}  The goal of maturity content prompt is to identify the maturity content indicator $MC = [MC_{1}, MC_{2}, ..., MC_{12}]$ for a given screenshot $s \in [s_{1},s_{2}, ..., s_{n} ]$. We design maturity content prompt $P_{\mathrm{MC}}$ according to the maturity rating policy as shown in Table \ref{Tab:ios_Rating_Policy}. As shown in Figure \ref{fig:cot_prompt}(a),  the maturity content prompt $P_{\mathrm{MC}}$ instructs the MLLM $f_{\mathrm{MLLM}}$ to output the  maturity content $MC$ as follows: 
\begin{equation}
    MC_s = f_{\mathrm{MLLM}}(MC| X_s, P_{\mathrm{MC}})
\end{equation}
where $X_s$ is the image of screenshot $s \in [s_{1},s_{2}, ..., s_{n} ]$. 

\iffalse
\begin{tcolorbox}
\setstretch{1.25}
Response to the maturity content (ID from 0 to 11) in the screenshot image based on the following options: \\
0. Cartoon or fantasy violence\\
1. Horror or fear-themed content \\
2. \\
3. \\
4. \\
5. \\
6. \\
7. \\
8. \\
9. \\
10. \\
11. 
\end{tcolorbox}
\fi

\textbf{Maturity Intensity Prompt:} 
After deriving maturity content $MC_s$ for screenshot $s$, we continue to evaluate the intensity of the identified maturity content.  As shown in Figure \ref{fig:cot_prompt}(b), the maturity intensity prompt $P_{\mathrm{MI}}$ instructs the MLLM model $f_{\mathrm{MLLM}}$ to output the  maturity intensity $I$ as follows:
\begin{equation}
    I_s = f_{\mathrm{MLLM}}(I| X_s, MC_s, P_{\mathrm{MI}}).
\end{equation}
For example, if we identified  ``Cartoon or Fantasy Violence" in a screenshot, then maturity intensity prompt $P_{\mathrm{MI}}$ is designed as follows: 
\begin{tcolorbox}
\setstretch{1.25}
Please response the maturity level of content: {\bf Cartoon or Fantasy Violence}, ranging from {\bf mild} to {\bf intense} in the screenshot image. 
\end{tcolorbox}

\textbf{Maturity Rate Prompt:} 
After identifying the maturity content and its intensity in each screenshot $s \in [s_{1},s_{2}, ..., s_{n} ]$, we rank them according to $\langle MC_s, I_s \rangle_{s \in \{1, ..., n\}}$ values, and obtain the top screenshot(s) $S^{\ast} \in \{s_{1},s_{2}, ..., s_{n}\}$. Then the maturity ratings $R$ are computed by combing the top screenshot(s) $S^{\ast}$  and the textual descriptions $D$.  As shown in Figure \ref{fig:cot_prompt}(c), the maturity rating prompt $P_{\mathrm{MR}}$ instructs the MLLM model $f_{\mathrm{MLLM}}$ to output the  maturity rating $R$ as follows:
\begin{equation}
    R = f_{\mathrm{MLLM}}(R| X_{s, s\in S^{\ast}},  D,  P_{\mathrm{MR}}).
\end{equation}
This prompt directs the MLLM to classify the app into one of four age-appropriate categories $\{4+, 9+, 12+, 17+\}$ based on the evaluated content. This prompt encourages the model to use its inherent knowledge to distill the extracted content themes and intensities into a clear age rating, employing a straightforward fusion strategy to assess the overall suitability for potential user groups.

\section{Experiments}
\label{sec:exp}

In this section, we present empirical evaluations to carefully assess our proposed method. 

\subsection{Experimental Setup}

{\noindent \bf{Dataset.}} We crawled data from App Store and curated a dataset for our empirical evaluations in this study.  The dataset consists of 1,281 mobile apps, categorized into four age groups: 324 apps rated 4+, 334 apps rated 9+, 394 apps rated 12+, and 229 apps rated 17+. Each app includes between 2 to 7 screenshot images along with a description. These elements serve as the multimodal inputs to the MLLM to assess the maturity level of the content. The apps were categorized across various genres to ensure diversity in content type (see Table \ref{Tab:ios_Rating_Policy}), which aids in generalizing the model's applicability.

{\noindent \bf{Baseline Models.}} App maturity rating is a multi-class classification problem, where we classify each app into one of the four age-appropriate categories $\{4+, 9+, 12+, 17+\}$.  To benchmark the performance of our proposed CoT-endowed method, we compare it against several established baseline models with different large language models. These models are selected based on their prior success in similar tasks such as text classification, image classification, and multimodal classification. 
\begin{itemize}
    \item Vicuna \cite{Vicuna} is an open-source chatbot trained by fine-tuning LLaMA on user-shared conversations and we use Vicuna to process app textual descriptions for maturity rating.
    \item GPT-3.5 and GPT-4 \cite{openai2024gpt4technicalreport}. We respectively use GPT-3.5 and GPT-4 to input app textual descriptions to return the maturity ratings.
    \item LLaVa-1.5 \cite{LLaVA-1.5} is a large multimodal model trained by combining a vision encoder and Vicuna for general-purpose visual and language understanding. We apply LLaVa-1.5 to process the app screenshots to return the maturity ratings. 
    \item GPT-4V \cite{openai2024gpt4technicalreport} is the Multimodal LLM developed by OpenAI by incorporating image modalities into large language models. We instruct GPT-4V to analyze single-modality (screenshot-only) and dual-modality (screenshot+description) to return the maturity ratings. 
\end{itemize}

 {\noindent \bf{Evaluation Metrics.}} Since app maturity rating is a multi-class classification problem,  we employ a variety of classification metrics to comprehensively evaluate the effectiveness of our proposed method in predicting app maturity levels. These metrics include accuracy, precision, recall, and F1-score; each provides insights into different aspects of model performance. Such metrics help us gauge not only the accuracy but also the consistency and reliability of the maturity ratings.

%This experimental setup, with its diverse dataset and robust comparative framework, ensures that our research is both comprehensive and applicable to real-world scenarios. The structured analysis of multimodal data through our CoT reasoning approach based on GPT-4V is anticipated to demonstrate significant advancements over baseline MLLM, thus providing deeper insights into the practical utility and scalability of our MLLM system. These evaluations are crucial for validating the model's efficacy in accurately assessing the maturity of mobile applications across varied content types.

\subsection{Comparison Against Baseline Methods}

\begin{table}[t]
\centering
\small
\caption{Performance comparisons between the proposed CoT-endowed MLLM framework for app maturity rating against the baseline approaches.}
\label{tab:performance_metrics}
\begin{tabular}{|c|c|c|c|c|c|}
\hline
\textbf{Method} & \textbf{Modality} & \textbf{Accuracy} & \textbf{Precision} & \textbf{Recall} & \textbf{F1-score} \\ \hline
Vicuna & Description-only & 40.98\% & 40.15\% & 41.47\% & 40.30\% \\ \hline
GPT-3.5 & Description-only & 56.60\% & 57.53\% & 56.20\% & 56.32\% \\ \hline
GPT-4 & Description-only & 64.87\% & 64.26\% & 65.20\% & 63.94\% \\ \hline
LLaVa-1.5 & Screenshot-only & 33.13\% & 40.24\% & 33.13\% & 29.38\% \\ \hline
GPT-4V & Screenshot-only & 61.67\% & 62.29\% & 61.16\% & 59.96\% \\ \hline
GPT-4V & Screenshot+Description & 70.34\% & 69.67\% & 71.49\% & 70.38\% \\ \hline
\textbf{CoT (Ours)} & Screenshot+Description & \textbf{71.82\%} & \textbf{71.29\%} & \textbf{73.17\%} & \textbf{72.00\%} \\ \hline
\end{tabular}
\end{table}

Table \ref{tab:performance_metrics} shows the results of our proposed  CoT-endowed MLLM framework for app maturity rating against the baseline approaches. Our approach consistently outperforms the baseline methods across all metrics with significant margins. 
Specifically, we first observe that different LLM models would lead to different performances. For example, for methods that only use textual description for maturity rating, GPT-4 outperforms GPT-3.5 and Vicuna; for methods that only use screenshots for maturity rating, GPT-4V outperforms LLaVa-1.5 (F1-score 59.96\% vs 29.38\%). Second, for the same LLM platform, the text-based method (GPT-4 with F1-score of 63.94\%) outperforms the image-based method (GPT-4V with F1-score of 59.96\%). Third, the multimodal method outperforms unimodal methods. By combining both screenshot and description, GPT-4V achieves an F1-score of 70.38\%, while the F1-score for the description-only method is 63.94\% and for the screenshot-only method is 59.96\%. Finally, our proposed approach achieves much more accurate maturity ratings than the best baseline that integrates multimodal data, including screenshots and descriptions (F1-score 72.00\% vs 70.38\%). The performance improvement shows that CoT-endowed prompting is much more effective for maturity rating reasoning through a sequence of reasoning steps.

\begin{figure}
\centering
\includegraphics[width=0.6\linewidth]{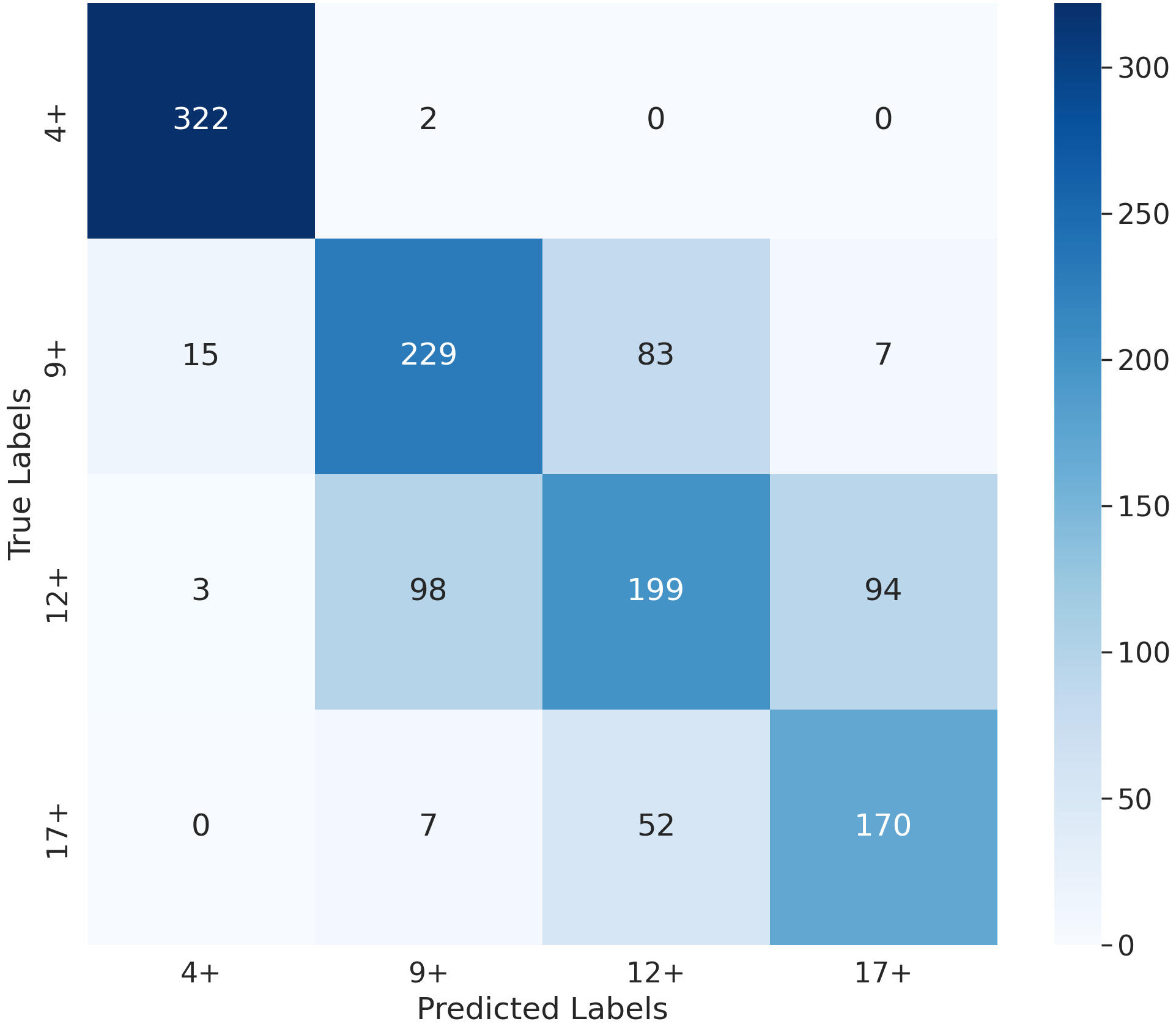}
%\vspace{-10pt}
\caption{Confusion matrix of the proposed CoT endowed MLLM framework for app maturity rating.}
\label{fig:confusion_matrix}
\end{figure}

We further study the performance of our proposed method in predicting different maturity levels. Figure \ref{fig:confusion_matrix} shows the confusion matrix of predicted versus actual age ratings, including 4+, 9+, 12+, and 17+.
Analyzing the matrix, we observe a high number of correct predictions (diagonal elements), indicating robust model accuracy across all age categories. 
Specifically, for the 4+ category, the framework achieved accurate predictions (precision of 94.7\% and recall of 99.38\%), which shows a strong ability to identify content suitable for all ages. 
In addition, for the 229  age 17+ apps, we observe  decrease in predictions (precision of 62.73.7\% and recall of 74.23\%). For example, 52 (22.7\%) of the age 17+ apps were predicted as 12+ app; 94 out of the 394 (23.85\%) age 12+ apps were classified as age 17+ apps. 
We observe similar confusion for classifying age 9+ apps and age 12+ apps. 
We speculate the following reasons that the performance decreases in classifying age 9+ apps, age 12+ apps, and age 17+ apps. First, maturity rating is a challenging task by nature. For example, according to rating policy, ``Infrequent/mild Mature or suggestive content" is categorized as age 9+ apps; meanwhile, ``Infrequent/mild Sexual content or nudity" is categorized as age 12+ apps. The boundary between these two concepts is not always clear, which makes it difficult to distinguish these two concepts by looking at app images and descriptions. Second, app developers might intentionally obscure mention of maturity contents in-app images and descriptions to promote their apps to more users. 

\subsection{Comparative Analysis of Different Multimodal Fusion Strategies for App Maturity Rating}

Typically, different multimodal fusion strategies would lead to different multimodal learning performances. This subsection evaluates the effectiveness of different fusion strategies in classifying app maturity, focusing on how they integrate multimodal information. We compare following multimodal fusion strategies. 

\begin{itemize}
\item \textbf{Basic Fusion Strategy:} This baseline method uses a single screenshot combined with the app description for maturity rating.

\item \textbf{Image-Caption Fusion Strategy:} This method involves using GPT-4V to create a caption for a selected screenshot, which is then combined with the app's main description to predict the maturity rate. This strategy leverages the descriptive power of GPT-4V to enhance textual understanding of visual content.

\item \textbf{Global CoT Fusion:} This strategy analyzes all available screenshots along with the description, offering a comprehensive assessment by considering all visual data for a complete understanding of the app's content.

\item \textbf{Selective CoT Fusion:} This approach combines textual description with the most indicative content through a selection of the top-ranked screenshot(s) for maturity rating.    
\end{itemize}

\begin{table}[t]
\centering
\small
\caption{Impacts of multimodal fusion strategies on app maturity rating.}
\label{tab:fushion_strategy}
\begin{tabular}{|c|c|c|c|c|}
\hline
\textbf{Fusion Method}  & \textbf{Accuracy} & \textbf{Precision} & \textbf{Recall} & \textbf{F1-score} \\ \hline
Basic Fusion   & 70.34\% & 69.67\% & 71.49\% & 70.38\% \\ \hline
Image-Caption Fusion  & 63.00\% & 63.22\% & 64.62\% & 63.24\% \\ \hline
Global CoT Fusion & 57.77\% & 58.42\% & 57.45\% & 56.23\% \\ \hline
%CoT-based & $n$ Images + Description & 58.42\% & 57.72\% & 56.87\% & 56.63\% \\ \hline
\textbf{Selective CoT Fusion}  & \textbf{71.82\%} & \textbf{71.31\%} & \textbf{73.17\%} & \textbf{72.02\%} \\ \hline
\end{tabular}
\end{table}
% $1$ Image + Description

Table \ref{tab:fushion_strategy} shows the classification performances of the aforementioned multimodal fusion strategies. We observe that the Selective CoT Fusion yields the best performance. This comparative analysis underscores the importance of strategic multimodal integration in achieving reliable maturity classification in mobile apps.

\subsection{Impact of Chain-of-Thought (CoT) Reasoning on Maturity Rating}

\begin{table}[t]
\centering
\small
\caption{Impacts of chain-of-thought (CoT) reasoning strategies on maturity rating.}
\label{tab:CoT_study}
\begin{tabular}{|c|c|c|c|c|}
\hline
 \textbf{CoT Scheme} & \textbf{Accuracy} & \textbf{Precision} & \textbf{Recall} & \textbf{F1-score} \\ \hline
 Text understanding + Image & 38.33\% & 58.04\% & 37.17\% & 32.62\% \\ \hline
Image selection + Text understanding & 41.30\% & 65.40\% & 39.51\% & 36.72\% \\ \hline
 \textbf{Image selection + Text (Ours)}  & \textbf{71.82\%} & \textbf{71.29\%} & \textbf{73.17\%} & \textbf{72.00\%} \\ \hline
\end{tabular}
\end{table}

Note that in our proposed  CoT endowed prompting for app maturity rating, we first identify and rank the screenshots according to the exist and intensity of maturity contents and then combine the top screenshot(s) and the textual
descriptions to determine the final maturity rating. 
To further evaluate the effectiveness of our proposed CoT on maturity rating, we conduct  experiments on different CoT reasoning strategies. 
Table \ref{tab:CoT_study} shows the performances of maturity rating with different CoT reasoning strategies. 
In the ``Text understanding + Image" CoT strategy, similar to the way to prompt MLLM to understand the exist and intensity of maturity content in an image, we  use the same method to comprehend app descriptions before we combine with the screenshot for final maturity rating. Accordingly, in the ``Image selection + Text understanding" CoT strategy, prompt MLLM to understand both images and text, and combine the results to access maturity levels. 
We can see that the performance of the ``Text understanding + Image" CoT strategy is notably poor, which in turn affects the overall effectiveness when combined with image selection-based CoT.
Specifically, the combination of CoT applied to both text and image modestly improves accuracy and precision but remains significantly lower than desired. Surprisingly, using CoT to identify and rank the screenshots and then combing text yields much better results.
These findings highlight the challenges of implementing CoT in text processing within this context and suggest that while CoT enhances image analysis, its application in the text needs further refinement to avoid diminishing the benefits of multimodal fusion strategies.

\begin{table}[t]
\centering
\small
\caption{Ablation analysis of chain-of-thought (CoT) reasoning on maturity rating.}
\label{tab:ablation_study}
\begin{tabular}{|c|c|c|c|c|}
\hline
 \textbf{CoT Scheme} & \textbf{Accuracy} & \textbf{Precision} & \textbf{Recall} & \textbf{F1-score} \\ \hline
 w/o CoT Image & 55.14\% & 59.72\% & 53.51\% & 53.58\% \\ \hline
 \textbf{w/ CoT Image  (Ours)}  & \textbf{71.82\%} & \textbf{71.29\%} & \textbf{73.17\%} & \textbf{72.00\%} \\ \hline
\end{tabular}
\end{table}

To further investigate the importance of the proposed CoT method, we conduct an ablation study by comparing two image selection methods for app maturity rating. 
Table \ref{tab:ablation_study} shows the results of the ablation study.  The ``w/ CoT Image" method applies CoT reasoning to systematically select the screenshot that is most indicative of the app's content maturity. Conversely, the ``w/o CoT Image" method randomly selects a screenshot from the remaining images that were not identified by the CoT approach. This design allows us to demonstrate the precision and effectiveness of the CoT methodology in enhancing the accuracy of maturity assessments by comparing it to a non-discriminative selection approach. Comparing the approach with and without CoT on images reveals that using CoT results in higher accuracy, precision, recall, and F1-score. This not only validates the efficacy of incorporating CoT but also emphasizes its critical role in improving the reliability and accuracy of automated content maturity assessments.

Through above two studies, we can see that our CoT approach leads to best results, emphasizing the need for its holistic application for accurate digital content classification.

\subsection{Case studies}
\subsubsection{Case Study 1: Evaluating Multimodal Fusion Strategies for Maturity Rating}
In this subsection, we present case studies from our experimental evaluations to illustrate the application of our MLLM for app maturity rating (see Table \ref{tab:case1}). These cases highlight how different fusion strategies compare against each other and the ground truth (GT) ratings. Each case study involves a specific mobile game featuring both visual and textual content, which our framework processes to predict the appropriate age rating.

\begin{table}
\centering
\small
\caption{Comparative performance across different fusion strategies. Basic Fusion ($P_{BF}$) uses a simple integration approach; Image-Caption Fusion ($P_{CF}$) adds descriptions generated from images; Global CoT Fusion ($P_{GloCoT}$) analyzes all screenshots; Selective CoT Fusion ($P_{SelCoT}$, our method) selects the most maturity rate  screenshot through CoT reasoning. Correct predictions are marked in \textcolor{blue}{blue}, and incorrect ones in \textcolor{red}{red}, illustrating each strategy's accuracy.}
\label{tab:case1}
\begin{tabular}{c|c|p{10cm}}
\hline
\textbf{No.} & \textbf{Image} & \textbf{Description} \\ %& \textbf{GT} & \textbf{P\_ours} & \textbf{P\_CF} & \textbf{P\_SF} \\
\hline
1 & \raisebox{-\totalheight}{\includegraphics[width=0.12\textwidth]{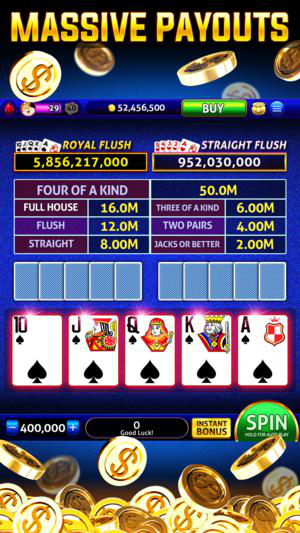}} & Join Club Vegas and become a Las Vegas VIP anytime, anywhere! Experience a new slot casino game with endless bonuses, huge jackpots, and the best free slot machines inspired by Vegas casinos. Claim your 1,000,000 COINS WELCOME BONUS and start spinning the reels of all the HOT video slots in this free casino game...  \\ 
\hline
& $GT$: 17 &  \textcolor{blue}{$P_{BF}$: 17} \hspace{0.5cm}  \textcolor{blue}{$P_{CF}$: 17}    \hspace{0.5cm}   \textcolor{blue}{$P_{GloCoT}$: 17}    \hspace{0.5cm} \textcolor{blue}{$P_{SelCoT}(ours)$: 17}\\ \hline
2                 &  \raisebox{-\totalheight} {\includegraphics[scale=0.14]{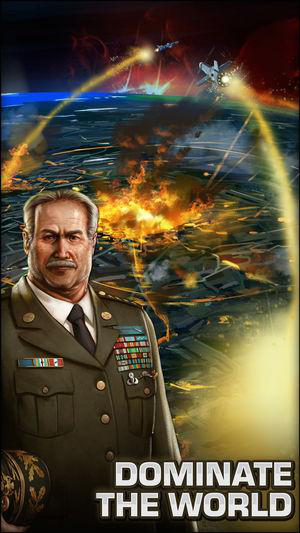}}                                
                  & free game in 26 countries and counting free app in 10 countries and counting Indulge your inner criminal mastermind in Crime City, brought to you by GREE—makers of hit games like Modern War!Breaking the rules is way more fun than playing by them! Build your criminal empire, one job at a time...         \\  \hline
& $GT$: 17 & \textcolor{blue}{$P_{BF}$: 17} \hspace{0.5cm}   \textcolor{blue}{$P_{CF}$: 17}    \hspace{0.5cm}    \textcolor{blue}{$P_{GloCoT}$: 17}  \hspace{0.5cm} \textcolor{blue}{$P_{SelCoT}(ours)$: 17}   \\ \hline
3                 &  \raisebox{-\totalheight} {\includegraphics[scale=0.14]{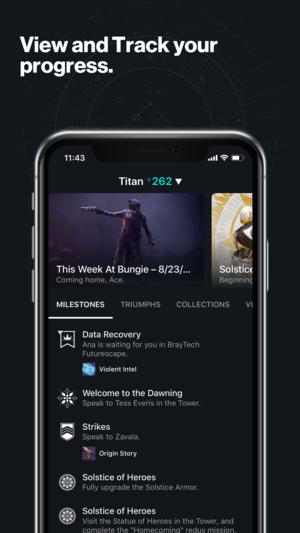}}                                
                  & The official Destiny 2 Companion App keeps you connected to your Destiny adventure wherever life takes you.   Join using PlayStation Network, Xbox Live, or Battle.net. COMPANION - See all the latest news and updates.  Discover what events and activities are live in the game.   View your progress towards your Triumphs and your Collections...    \\  \hline
& $GT$: 9 & \textcolor{red}{$P_{BF}$: 12} \hspace{0.5cm}  \textcolor{red}{$P_{CF}$: 12}    \hspace{0.5cm}   \textcolor{red}{$P_{GloCoT}$: 12}  \hspace{0.5cm} \textcolor{red}{$P_{SelCoT}(ours)$: 12}   \\ \hline
4                 &  \raisebox{-\totalheight} {\includegraphics[scale=0.14]{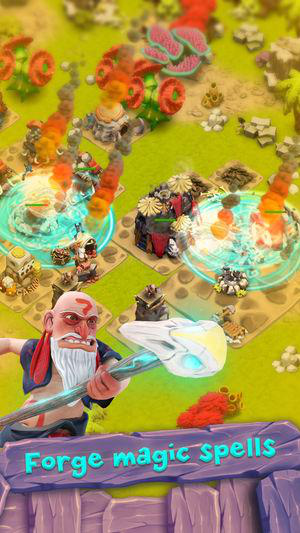}}                               
                  &   Fend off dangers of the wild, battle ancient armies and create alliances in your mission to conquer the world. As village chief, you are to build a prosperous settlement while defending your people from prehistoric dangers. This will be no easy task for other clans occupy this world; clans whose intentions might not be so pure...   \\  \hline
& $GT$: 9 &  \textcolor{red}{$P_{BF}$: 12}  \hspace{0.5cm}  \textcolor{red}{$P_{CF}$: 12}    \hspace{0.5cm}    \textcolor{red}{$P_{GloCoT}$: 12}  \hspace{0.5cm} \textcolor{blue}{$P_{SelCoT}(ours)$: 9}  \\ 
\hline
\end{tabular}
\vspace{-15pt}
\end{table}

Table \ref{tab:case1} indicates that for an app suitable for older teenagers (17+) featuring casino-related content or realistic violence, all methods—Basic Fusion ($P_{BF}$), Image-Caption Fusion ($P_{CF}$), Global CoT ($P_{GloCoT}$), and our approach ($P_{SelCoT}(ours)$) —uniformly rated this app as 17+, which matches the ground truth (refer to cases No.1 and No.2). 
Case No.3 demonstrates that although the GT rates this app as suitable for ages 9+, all models predicted a rating of 12+. This discrepancy may arise because the image lacks clear information related to age-appropriate content, leading to a prediction bias in every fusion strategy employed.
Case No.4 indicates that the GT recommends a 9+ rating, appropriate for children able to understand strategic game mechanics. However, our model accurately predicted this rating, while the other models overestimated the age suitability at 12+.

These case studies demonstrate the varied capabilities of different fusion strategies in our multimodal LLM framework. Our method shows a consistent alignment with ground truth ratings, emphasizing its robustness in accurately assessing app content through detailed analysis of both text and imagery. The discrepancies observed in other models highlight the challenges in multimodal content evaluation, especially in interpreting the subtleties of visual elements and narrative context. This comparative analysis not only validates our model's effectiveness but also illustrates the potential for further refinement in automated maturity rating systems.

\subsubsection{Case Study 2: Comparative Analysis of CoT-based method and Non-CoT-based method}
Some case studies further explore the impact of selected images on the performance of our maturity evaluation framework (see Table \ref{tab:case2}). Each case contrasts the effect of using different images from the same app on the predicted maturity ratings, examining the role of CoT in this process.

\begin{table}[t]
    \centering
    \small
    \caption{Comparative Analysis of Image Selection and Maturity Rating Prediction. Correct predictions are highlighted in \textcolor{blue}{blue}, and incorrect ones in \textcolor{red}{red}, to visually represent the accuracy of each method.}
    \label{tab:case2}
    \begin{tabular}{c|c|c}
        \hline
        No. & Selected Image & Other Images \\

        \hline
        1 & \includegraphics[height=1.6cm]{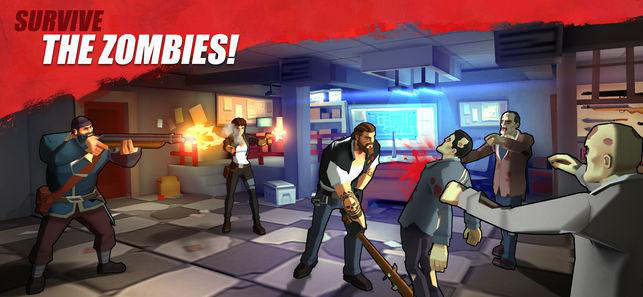} & \includegraphics[height=1.6cm]{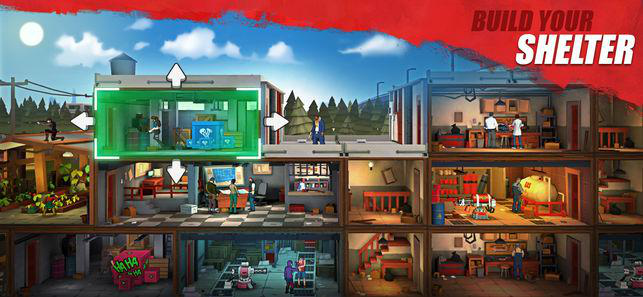} \includegraphics[height=1.6cm]{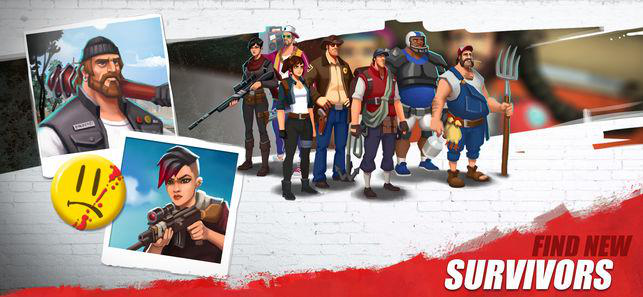} \\
        \hline
        GT: 17 & \textcolor{red}{w/ CoT: 12} & \textcolor{red}{w/o CoT: 12} \\
        \hline
        2 & \includegraphics[height=2cm]{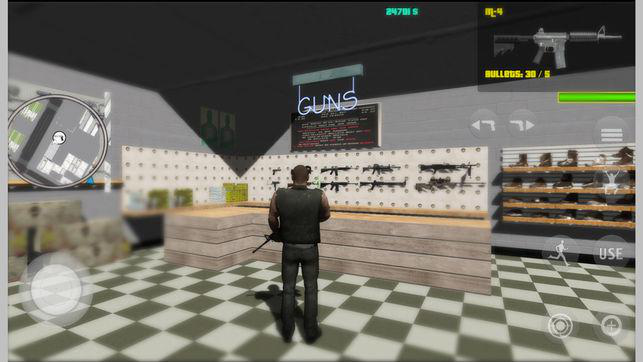} & \includegraphics[height=2cm]{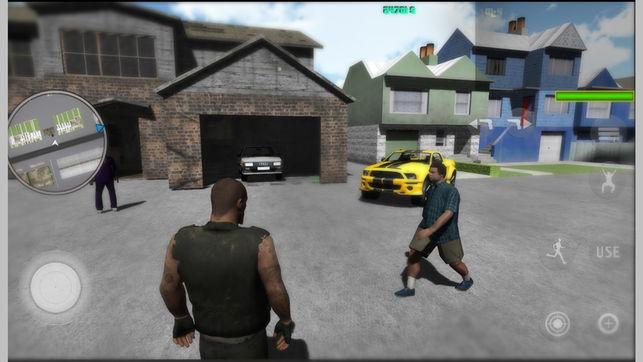} \includegraphics[height=2cm]{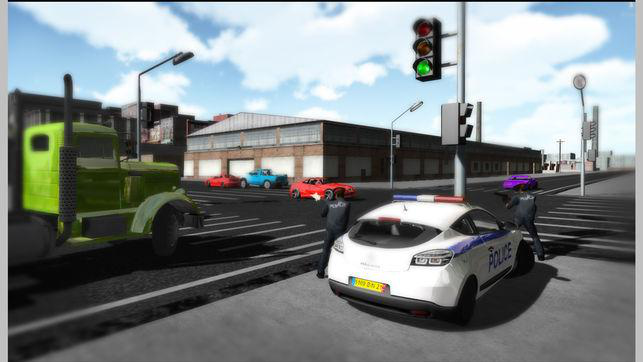}\\
        \hline
        GT: 17 & \textcolor{blue}{w/ CoT: 17} & \textcolor{red}{w/o CoT: 12} \\
        \hline
        3 & \includegraphics[height=2cm]{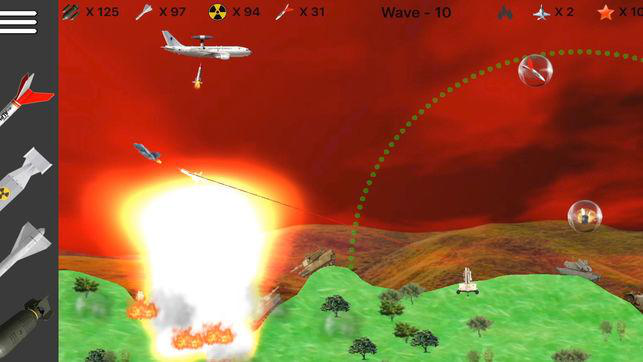} & \includegraphics[height=2cm]{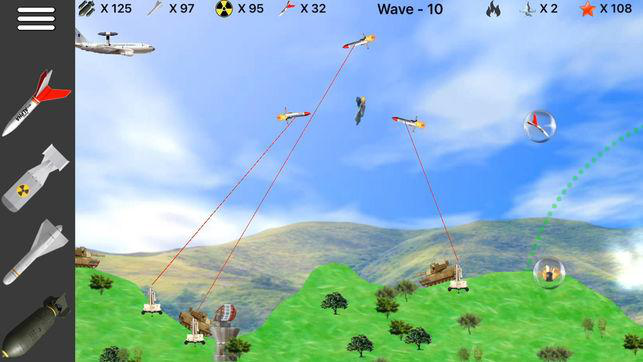} \includegraphics[height=2cm]{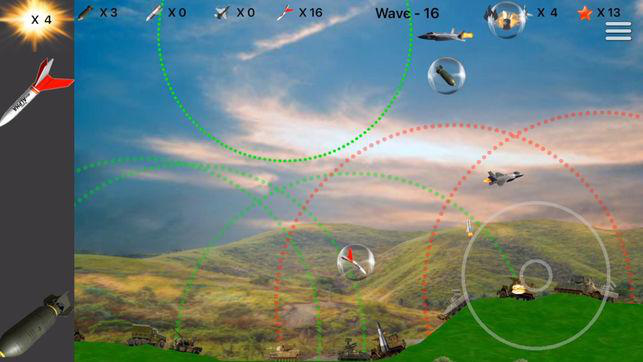} \\
        \hline
        GT: 9 & \textcolor{red}{w/ CoT: 17} & \textcolor{red}{w/o CoT: 12} \\
        \hline
               4 & \includegraphics[height=2cm]{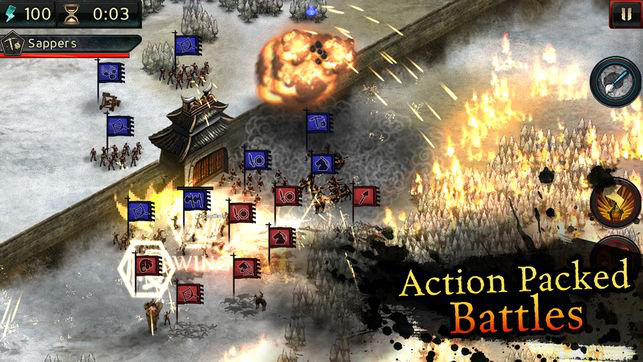} & \includegraphics[height=2cm]{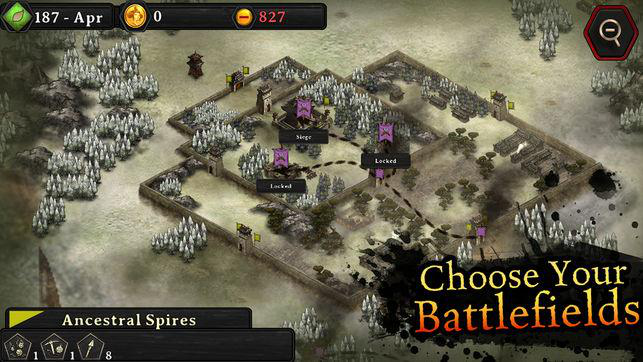} \includegraphics[height=2cm]{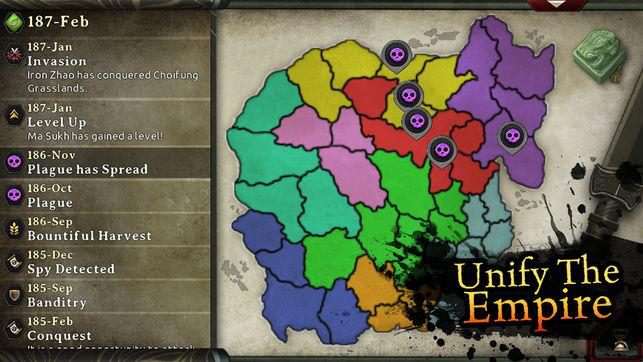} \\
        \hline

        GT: 9 & \textcolor{red}{w/ CoT: 12} & \textcolor{blue}{w/o CoT: 9} \\
                        \hline
    \end{tabular}
    \vspace{-10pt}
\end{table}

Case 1 in Table \ref{tab:case2} shows that the GT rating for this game is 17, likely due to its horror and violent themes. Interestingly, both with and without the use of CoT, the model rated the app as suitable for 12+. This discrepancy suggests that while the selected images were significant, they did not fully capture the maturity level expected by the GT, indicating a need for more comprehensive visual analysis or an adjustment in how images are weighted by the model.
Case 2 demonstrates the effectiveness of CoT in interpreting more complex or nuanced content accurately when presented with more explicit imagery.
Case 3 shows that the GT for this app is surprisingly low at 9, despite the model with CoT assigning a rating of 17, influenced perhaps by the aggressive thematic content of the selected image. The rating without CoT resulted in a more moderate 12, indicating sensitivity to image selection and the challenge of aligning model perception with community standards.
Case 4 presents an interesting reversal where the model with CoT suggested a 12+ rating, while without CoT, the rating matched the GT at 9. This suggests that CoT might be emphasizing the warfare aspect more heavily, impacting the perceived suitability for younger players.

These case studies underline the critical role of image selection in automated content rating systems and highlight how different images can lead to varying interpretations of app content. They also emphasize the importance of implementing robust reasoning mechanisms like CoT, which can discern subtleties in content that might otherwise lead to underestimation or overestimation of maturity levels. This analysis is crucial for refining the accuracy of our model, ensuring that it not only performs well across a broad spectrum of apps but also aligns closely with established content rating standards.

\subsection{Discussion}

Multimodal large language models (MLLMs) represented by GPT-4V have shown impressive performances in different tasks such as image and text understanding. As shown in our study, applying  MLLMs to a real-world application still requires careful consideration. For example, in the maturity rating task in this study,  CoT reasoning can improve the performance further than the direct application of GPT-4V. Also, the effectiveness of these fusion strategies can vary based on the type of content and the specific requirements of the application. For example, the analysis of Basic Fusion, Image-Caption Fusion, and Global CoT Fusion strategies highlights the varying degrees of success and challenges, suggesting that no one-size-fits-all solution exists. This insight is crucial for developers and researchers looking to optimize content classification systems for specific user needs and content types.

Additionally, the exploration of more advanced multimodal learning techniques that seamlessly integrate various data types (text, image, video, etc.) could further refine the effectiveness of maturity rating systems. Such advancements could pave the way for more nuanced and context-aware AI systems capable of adapting to the ever-evolving landscape of digital content. These future directions not only promise to enhance the technical capabilities of maturity assessment models but also contribute to safer and more appropriate digital environments for diverse user groups.

\section{Conclusion and Future Work} 

In this work, we presented the first systematic study on app maturity rating with multimodal large language models (MLLMs).
We designed chain-of-thought (CoT) endowed prompting to instruct MLLM to better understand the images and text associated with apps step by step to determine maturity levels for apps. Our comprehensive experiments have demonstrated that the integration of CoT with multimodal data significantly improves the accuracy and reliability of maturity ratings. The CoT methodology, particularly when applied to a combination of textual and visual data, provides a deeper understanding of content, enabling more precise classifications and proving superior to traditional approaches.

\emph{Limitations and Future Research.} There are several limitations in this work and some interesting future research directions. First, due to the limit on  GPT-4V access, we evaluated our method on a relatively small dataset. A better model evaluation can be achieved by experiments with more data. 
Second, more advanced methods could be applied to improve the maturity rating performances. 
Finally, the implications of this study are vast, suggesting avenues for further enhancements in AI-driven content evaluation systems. The success of the Selective CoT Fusion strategy in our tests highlights the potential for targeted data processing techniques to refine AI assessments in various digital environments. As digital content continues to grow in complexity and volume, the need for advanced, nuanced, and reliable content filtering technologies becomes increasingly critical. Our research contributes to this ongoing development, laying the groundwork for future innovations that will continue to enhance safe and appropriate content consumption across digital platforms.

%\section{ACKNOWLEDGEMENTS}
%This work is partially supported by the NSF under grants IIS-2107172, IIS-2140785, CNS-1940859, CNS-1814825, IIS-2027127, IIS-2040144, IIS-1951504,  OAC-1940855, and NIJ 2018-75-CX-0032.

%\section{CONTRIBUTORS}
%CH designed the study, collected the data, built the model, and ran the experiment. BL helped design the study. This study was conducted under the supervision of YY and XL. XL provided substantial contributions to paper writing and submission.

\singlespacing
 \bibliographystyle{elsarticle-num} 
 \bibliography{cas-refs}

\begin{thebibliography}{10}
\expandafter\ifx\csname url\endcsname\relax
  \def\url#1{\texttt{#1}}\fi
\expandafter\ifx\csname urlprefix\endcsname\relax\def\urlprefix{URL }\fi
\expandafter\ifx\csname href\endcsname\relax
  \def\href#1#2{#2} \def\path#1{#1}\fi

\bibitem{google_play_stats_2024}
{Business of Apps}, \href{https://www.businessofapps.com/data/google-play-statistics/}{Google play store statistics} (2024).
\newline\urlprefix\url{https://www.businessofapps.com/data/google-play-statistics/}

\bibitem{apple_store_stats_2024}
{Business of Apps}, \href{https://www.businessofapps.com/data/app-stores/}{Apple app store statistics} (2024).
\newline\urlprefix\url{https://www.businessofapps.com/data/app-stores/}

\bibitem{anderson2001effects}
C.~A. Anderson, B.~J. Bushman, Effects of violent video games on aggressive behavior, aggressive cognition, aggressive affect, physiological arousal, and prosocial behavior: A meta-analytic review of the scientific literature, Psychological science 12~(5) (2001) 353--359.

\bibitem{hill2016media}
D.~Hill, N.~Ameenuddin, Y.~L. Reid~Chassiakos, C.~Cross, J.~Hutchinson, A.~Levine, R.~Boyd, R.~Mendelson, M.~Moreno, W.~S. Swanson, et~al., Media and young minds, Pediatrics 138~(5) (2016).

\bibitem{chen2013app}
Y.~Chen, H.~Xu, Y.~Zhou, S.~Zhu, Is this app safe for children? a comparison study of maturity ratings on android and ios applications, in: Proceedings of the 22nd international conference on World Wide Web, 2013, pp. 201--212.

\bibitem{hu2015protecting}
B.~Hu, B.~Liu, N.~Z. Gong, D.~Kong, H.~Jin, Protecting your children from inappropriate content in mobile apps: An automatic maturity rating framework, in: Proceedings of the 24th ACM International on Conference on Information and Knowledge Management, 2015, pp. 1111--1120.

\bibitem{zhou2022automatic}
C.~Zhou, X.~Zhan, L.~Li, Y.~Liu, Automatic maturity rating for android apps, in: Proceedings of the 13th Asia-Pacific Symposium on Internetware, 2022, pp. 16--27.

\bibitem{baltruvsaitis2018multimodal}
T.~Baltrušaitis, C.~Ahuja, L.-P. Morency, Multimodal machine learning: A survey and taxonomy, IEEE Transactions on Pattern Analysis and Machine Intelligence 41~(2) (2018) 423--443.

\bibitem{ngiam2011multimodal}
J.~Ngiam, A.~Khosla, M.~Kim, J.~Nam, H.~Lee, A.~Y. Ng, Multimodal deep learning, in: Proceedings of the 28th international conference on machine learning (ICML-11), 2011, pp. 689--696.

\bibitem{hu2021detection}
C.~Hu, M.~Yin, B.~Liu, X.~Li, Y.~Ye, Detection of illicit drug trafficking events on instagram: A deep multimodal multilabel learning approach, in: CIKM, 2021, pp. 3838--3846.

\bibitem{hu2021identifying}
C.~Hu, M.~Yin, B.~Liu, X.~Li, Y.~Ye, Identifying illicit drug dealers on instagram with large-scale multimodal data fusion, ACM Transactions on Intelligent Systems and Technology (TIST) 12~(5) (2021) 1--23.

\bibitem{hu2023fine}
C.~Hu, B.~Liu, Y.~Ye, X.~Li, Fine-grained classification of drug trafficking based on instagram hashtags, Decision Support Systems 165 (2023) 113896.

\bibitem{ramachandram2017deep}
D.~Ramachandram, G.~W. Taylor, Deep multimodal learning: A survey on recent advances and trends, IEEE signal processing magazine 34~(6) (2017) 96--108.

\bibitem{wei2022emergent}
J.~Wei, Y.~Tay, R.~Bommasani, C.~Raffel, B.~Zoph, S.~Borgeaud, D.~Yogatama, M.~Bosma, D.~Zhou, D.~Metzler, et~al., Emergent abilities of large language models, arXiv preprint arXiv:2206.07682 (2022).

\bibitem{chang2024survey}
Y.~Chang, X.~Wang, J.~Wang, Y.~Wu, L.~Yang, K.~Zhu, H.~Chen, X.~Yi, C.~Wang, Y.~Wang, et~al., A survey on evaluation of large language models, ACM Transactions on Intelligent Systems and Technology 15~(3) (2024) 1--45.

\bibitem{zhao2023survey}
W.~X. Zhao, K.~Zhou, J.~Li, T.~Tang, X.~Wang, Y.~Hou, Y.~Min, B.~Zhang, J.~Zhang, Z.~Dong, et~al., A survey of large language models, arXiv preprint arXiv:2303.18223 (2023).

\bibitem{hu2023unveiling}
C.~Hu, B.~Liu, X.~Li, Y.~Ye, Unveiling the potential of knowledge-prompted chatgpt for enhancing drug trafficking detection on social media, arXiv preprint arXiv:2307.03699 (2023).

\bibitem{openai2023gpt4}
OpenAI, \href{https://openai.com/research/gpt-4}{Gpt-4: Advancements in multimodality} (2023).
\newline\urlprefix\url{https://openai.com/research/gpt-4}

\bibitem{schramm2024impact}
S.~Schramm, S.~Preis, M.-C. Metz, K.~Jung, B.~Schmitz-Koep, C.~Zimmer, B.~Wiestler, D.~M. Hedderich, S.~H. Kim, Impact of multimodal prompt elements on diagnostic performance of gpt-4 (v) in challenging brain mri cases, medRxiv (2024) 2024--03.

\bibitem{pillai2024evaluating}
A.~Pillai, S.~Parappally-Joseph, J.~Hardin, Evaluating the diagnostic and treatment recommendation capabilities of gpt-4 vision in dermatology, medRxiv (2024) 2024--01.

\bibitem{jia2024can}
S.~Jia, R.~Lyu, K.~Zhao, Y.~Chen, Z.~Yan, Y.~Ju, C.~Hu, X.~Li, B.~Wu, S.~Lyu, Can chatgpt detect deepfakes? a study of using multimodal large language models for media forensics, in: Proceedings of the IEEE/CVF Conference on Computer Vision and Pattern Recognition, 2024, pp. 4324--4333.

\bibitem{lian2024gpt}
Z.~Lian, L.~Sun, H.~Sun, K.~Chen, Z.~Wen, H.~Gu, B.~Liu, J.~Tao, Gpt-4v with emotion: A zero-shot benchmark for generalized emotion recognition, Information Fusion 108 (2024) 102367.

\bibitem{ahn2024large}
J.~Ahn, R.~Verma, R.~Lou, D.~Liu, R.~Zhang, W.~Yin, Large language models for mathematical reasoning: Progresses and challenges, arXiv preprint arXiv:2402.00157 (2024).

\bibitem{wadhawan2024contextual}
R.~Wadhawan, H.~Bansal, K.-W. Chang, N.~Peng, Contextual: Evaluating context-sensitive text-rich visual reasoning in large multimodal models, arXiv preprint arXiv:2401.13311 (2024).

\bibitem{wei2022chain}
J.~Wei, X.~Wang, D.~Schuurmans, M.~Bosma, F.~Xia, E.~Chi, Q.~V. Le, D.~Zhou, et~al., Chain-of-thought prompting elicits reasoning in large language models, Advances in neural information processing systems 35 (2022) 24824--24837.

\bibitem{kojima2022large}
T.~Kojima, S.~S. Gu, M.~Reid, Y.~Matsuo, Y.~Iwasawa, Large language models are zero-shot reasoners, Advances in neural information processing systems 35 (2022) 22199--22213.

\bibitem{wang2022self}
X.~Wang, J.~Wei, D.~Schuurmans, Q.~Le, E.~Chi, S.~Narang, A.~Chowdhery, D.~Zhou, Self-consistency improves chain of thought reasoning in language models, arXiv preprint arXiv:2203.11171 (2022).

\bibitem{zhou2022least}
D.~Zhou, N.~Sch{\"a}rli, L.~Hou, J.~Wei, N.~Scales, X.~Wang, D.~Schuurmans, C.~Cui, O.~Bousquet, Q.~Le, et~al., Least-to-most prompting enables complex reasoning in large language models, arXiv preprint arXiv:2205.10625 (2022).

\bibitem{Vicuna}
{Vicuna}, \href{https://lmsys.org/blog/2023-03-30-vicuna/}{Vicuna: An open-source chatbot impressing gpt-4 with 90\%\* chatgpt quality} (2023).
\newline\urlprefix\url{https://lmsys.org/blog/2023-03-30-vicuna/}

\bibitem{openai2024gpt4technicalreport}
OpenAI, \href{https://arxiv.org/abs/2303.08774}{Gpt-4 technical report} (2024).
\newblock \href {http://arxiv.org/abs/2303.08774} {\path{arXiv:2303.08774}}.
\newline\urlprefix\url{https://arxiv.org/abs/2303.08774}

\bibitem{LLaVA-1.5}
H.~Liu, C.~Li, Y.~Li, Y.~J. Lee, \href{https://arxiv.org/abs/2310.03744}{Improved baselines with visual instruction tuning} (2024).
\newblock \href {http://arxiv.org/abs/2310.03744} {\path{arXiv:2310.03744}}.
\newline\urlprefix\url{https://arxiv.org/abs/2310.03744}

\end{thebibliography}





% \end{thebibliography}
\end{document}